# Low-dose 4DCT Reconstruction via Temporal Non-local Means[†]

**Zhen Tian**
Department of Biomedical Engineering, Graduate School at Shenzhen, Tsinghua University, Shenzhen, 518055, China, and Center for Advanced Radiotherapy Technologies, University of California San Diego, La Jolla, CA 92093-0843, USA

**Xun Jia**
Center for Advanced Radiotherapy Technologies and Department of Radiation Oncology, University of California San Diego, La Jolla, CA 92093-0843, USA

**Bin Dong**
Department of Mathematics, University of California San Diego, La Jolla, CA 93093-0112, USA

**Yifei Lou**
Department of Mathematics, University of California Los Angeles, Los Angeles, CA 90095

**Steve B. Jiang**[a]
Center for Advanced Radiotherapy Technologies and Department of Radiation Oncology, University of California San Diego, La Jolla, CA 92093-0843, USA

**Purpose**: Four-dimensional computed tomography (4DCT) has been widely used in cancer radiotherapy for accurate target delineation and motion measurement for tumors in thorax and upper abdomen areas. However, its prolonged scanning duration causes a considerably increase of radiation dose compared with the conventional CT, which is a major concern in its clinical application. This work is to develop a new algorithm to reconstruct 4DCT images from undersampled projections acquired at low mAs levels in order to reduce the imaging dose.

**Methods:** Conventionally, each phase of 4DCT is reconstructed independently using the filtered backprojection (FBP) algorithm. The basic idea of our new algorithm is that, by utilizing the common information among different phases, the input information required to reconstruct image of high quality, and thus the imaging dose, can be reduced. We proposed a temporal non-local means (TNLM) method to explore the inter-phase similarity. All phases of the 4DCT images are reconstructed simultaneously by minimizing a cost function consisting of a data fidelity term and a TNLM regularization term. We utilized a modified forward-

---

backward splitting algorithm and a Gauss-Jacobi iteration method to efficiently solve the minimization problem. The algorithm was also implemented on graphics processing unit (GPU) to improve the computational speed. Our reconstruction algorithm has been tested on a digital NCAT thorax phantom in three low dose scenarios: all projections with low mAs level, undersampled projections with high mAs level and undersampled projections with low mAs level.

**Results:** In all three low dose scenarios, our new algorithm generates visually much better CT images containing less image noise and streaking artifacts compared with the standard FBP algorithm. Quantitative analysis shows that, by comparing our TNLM algorithm with the standard FBP algorithm, the contrast-to-noise ratio has been improved by a factor of 3.9-10.2 and the signal-to-noise ratio has been improved by a factor of 2.1-5.9, depending on the cases. In the situation of undersampled projection data, the majority of the streaks in the images reconstructed by FBP can be suppressed using our algorithm. The total reconstruction time for all 10 phases of a slice ranges from 40 to 90 seconds on an NVIDIA Tesla C1060 GPU card.

**Conclusions:** The experimental results indicate that our new algorithm outperforms the conventional FBP algorithm in effectively reducing the image artifacts due to undersampling and suppressing the image noise due to the low mAs level.

**1. Introduction**

    Four-dimensional computed tomography (4DCT) has been widely used for treatment simulation in radiotherapy of tumors with respiratory motion[1-6], in that it can provide time-resolved volumetric images. Currently, there are two different methods for 4DCT acquisition and sorting, namely retrospective slice sorting[1-6] and prospective sinogram selection[7]. For the retrospective slice sorting method, the projection data are continuously acquired at each couch position for a time interval slightly longer than a full respiratory cycle, either in cine mode or in helical mode with a very low pitch[1-6]. Multiple slices corresponding to different acquisition time points are reconstructed at each couch position and then sorted into respiratory phase bins using various respiratory signals[1-6, 8-10]. For prospective sinogram selection method, the CT scanner is triggered by the respiratory signal for projection data acquisition[7]. The projection data within the same phase bin are used to reconstruct CT slices corresponding to that breathing phase.

    No matter which method is used, the reconstructed CT slices at different breathing phases represent a set of time-resolved volumetric images, which called 4DCT images and can resolve the organ motions and reduce motion artifacts to a satisfactory extent[11]. However, the prolonged acquisition time results in a considerably increased radiation dose. For example, given the standard 4DCT parameters (140kVp, 95mA, 0.5s per rotation), the radiation dose of a 4DCT scan is about 6 times of that of a typical helical CT scan[12]. This fact has become a major concern in the clinical application of 4DCT and thus it is highly desirable to reduce its imaging dose.

    Intuitively, the radiation dose can be reduced by lowering the mAs level and/or decreasing the number of acquired projections[†]. However, these approaches in general will lead to amplified image noise and obvious streaking artifacts in the reconstructed 4DCT images, if conventional filtered backprojection (FBP) algorithm is used. In the current FBP-based 4DCT reconstruction algorithm, each phase of the 4DCT images is reconstructed independently based on the acquired projection data associated with it. This method, though simple, completely neglected the highly temporal correlation between 4DCT images at successive phases, as apparently same anatomical features exist in successive phases of 4DCT images with slight motion and deformation. It is expected that taking into account this information during reconstruction, one can reconstruct 4DCT image with high quality, even at the low dose contexts of low mAs level and/or undersampled projection data.

    Inspired by this idea, in this paper we propose a new 4DCT reconstruction algorithm by utilizing the aforementioned temporal correlations between images at successive phases. In our algorithm, CT slices corresponding to different phases are reconstructed simultaneously as opposed to independently in conventional FBP-type algorithms. In particular, the temporal regularization is imposed between successive phases via a Temporal Non-local Means (TNLM) term to take the inter-phase correlation into account. Our idea of the TNLM function is inspired by the so called Non-local Means (NLM) method[16] originated in image processing

---

[†] Decreasing the number of projections is not straightforward on currently available commercial CT scanners due to the use of continuous x-ray generation mode. However, technically it is possible to modify the scanners to operate in high-frequency pulsed mode[13-15], if there is a clinical need (such as the one suggested in this paper).

field. The NLM method assumes that there are lots of repetitive structures contained in an image and thus utilizes the similar image features at different spatial locations in the same image to constructively enhance each other[17]. This assumption, though valid in many cases in image processing problems, it may not hold in medical images. However, it is reasonable to believe that along temporal direction there exist repetitive features in the 4DCT images. Thus we propose the novel TNLM method which extends the original NLM method into the temporal domain by exploring the similarity between neighboring phases in the context of 4DCT reconstruction.

The rest of this paper is organized as following. In Section 2 we will present the new 4DCT reconstruction method. The experimental results with this new method will be given in Section 3. We then give discussion and conclusions in Section 4.

## 2. Methods

*2.1 Reconstruction model*

Let us divide a respiratory cycle into $N$ phases labeled by $i = 1, 2, \ldots, N$. The 4DCT image of phase $i$ is denoted by a vector $f_i$. $P_i$ is the projection matrix of phase $i$ that maps the image $f_i$ into a set of projections corresponding to various projection angles. The measured set of projections is denoted by a vector $y_i$. We attempt to reconstruct the 4DCT images by solving the following optimization problem:

$$\{f_i\} = \text{argmin}_{\{f_i\}} \sum_{i=1}^{N} \left\{ (P_i f_i - y_i)^T \Sigma^{-1} (P_i f_i - y_i) + \frac{\mu}{2} [J(f_i, f_{i-1}) + J(f_i, f_{i+1})] \right\}, \quad (1)$$

where the first term in the summation is a data fidelity term, ensuring that the projections of the reconstructed 4DCT image at each phase matches the corresponding observed projections. The symbol $T$ denotes the matrix transpose. The covariance matrix $\Sigma$ is a diagonal matrix with its non-zero elements corresponding to the variance of the pixel values of the measured projection images[18]. The second term in Eq. (1), $J(\cdot,\cdot)$, is the regularization term, and the parameter $\mu$ adjusts the relative weight between the data fidelity term and the regularization term.

In this paper, we propose a new TNLM function as the temporal regularization imposed on neighboring phases to explore the inter-phase similarity. A periodic boundary condition along the temporal direction is assumed, *e.g.*, $f_{N+1} = f_1$. For two images at different phases, $f_i$ and $f_j$, $J(f_i, f_j)$ is defined as

$$J(f_i, f_j) = \sum_x \sum_y [f_i(x) - f_j(y)]^2 \omega_{i,j}^*(x, y). \quad (2)$$

The weighting factors $\omega_{i,j}^*(x, y)$ are independent of $f_i(x)$ and $f_j(x)$ but defined according to the ground truth images $f_i^*(x)$ and $f_j^*(x)$ as

$$\omega_{i,j}^*(x, y) = \frac{1}{Z} \exp\left[-\frac{\|R_i^*(x) - R_j^*(y)\|_2^2}{h^2}\right], \quad (3)$$

where $R_i^*(x)$ denotes a square patch on the image $f_i^*$ centering at a pixel $x$ and $\|R_i^*(x) - R_j^*(y)\|_2^2$ is the L$_2$-norm of the difference between $R_i^*(x)$ and $R_j^*(y)$. $Z$ is a normalization parameter such that $\sum_y \omega_{i,j}^*(x,y) = 1$. $h$ is a parameter that adjusts to what extent we would like to enforce the similarity between patches.

## 2.2 Optimization Approach

To solve the optimization problem in Eq.(1), we implement a forward-backward splitting[19, 20], where the solution to Eq.(1) can be obtained by alternatively performing the following two steps till convergence:

$$g_i^{(k)} = f_i^{(k-1)} - \frac{1}{\beta} P_i^T \Sigma^{-1} (P_i f_i^{(k-1)} - y_i), \forall i \tag{4}$$

$$\begin{aligned} \{f_i^{(k)}\} &= \mathrm{argmin}_{\{f_i\}} E_1[\{f_i\}] \\ &= \mathrm{argmin}_{\{f_i\}} \sum_{i=1}^N \left\{ \left\| f_i - g_i^{(k)} \right\|_2^2 + \frac{\mu}{2\beta} [J(f_i, f_{i-1}) + J(f_i, f_{i+1})] \right\}, \end{aligned} \tag{5}$$

where the superscript $k$ is the index for iteration steps. $g_i$ is auxiliary vector and $\beta > 0$ is a constant introduced by the splitting algorithm. Note that Eq.(4) is actually one step of gradient descent algorithm towards a problem minimizing an energy function. $E_2[f_i] = (P_i f_i - y_i)^T \Sigma^{-1} (P_i f_i - y_i)$. The introduction of $\beta$ actually controls the step size of the gradient descent algorithm for numerical stability purpose. In practice, it is found that by substituting this one step gradient descent with a conjugate gradient least square (CGLS) method[21] for the minimization of $E_2[f_i]$, the overall convergence can be enhanced, although the convergence after this modification is not mathematically proven. The CGLS algorithm by itself is an iterative algorithm. In each iteration $k$, we use the images obtained from the last iteration, *i.e.* $f_i^{(k-1)}$ as an initial guess. Let us denote $P_{ni} = \Sigma^{-1/2} P_i$ and $y_{ni} = \Sigma^{-1/2} y_i$. The detailed implementation of this CGLS algorithm is performed as follows:

---
CGLS Algorithm :

Initialinize: $m = 0$, $u_i^{(0)} = f_i^{(k-1)}$, $r_i^{(0)} = y_{ni} - P_{ni} u_i^{(0)}$, $s_i^{(0)} = P_{ni}^T r_i^{(0)}$. Do the Steps 1-5 for $M_1$ times.
1. $a_i^{(m)} = \left\| P_{ni}^T r_i^{(m)} \right\|_2^2 / \left\| P_{ni} s_i^{(m)} \right\|_2^2$
2. $u_i^{(m+1)} = u_i^{(m)} + a_i^{(m)} s_i^{(m)}$
3. $r_i^{(m+1)} = r_i^{(m)} - a_i^{(m)} P_{ni} s_i^{(m)}$
4. $b_i^{(m)} = \left\| P_{ni}^T r_i^{(m+1)} \right\|_2^2 / \left\| P_{ni}^T r_i^{(m)} \right\|_2^2$
5. $s_i^{(m+1)} = P_{ni}^T r_i^{(m+1)} + b_i^{(m)} s_i^{(m)}$
6. $g_i^{(k)} = u_i^{(M_1)}$

---

Here the superscript $m$ is the iteration step for the CGLS algorithm. In practice, it is not necessary to carry out this CGLS algorithm very precisely in each outer loop $k$, since the purpose of this CGLS step is only to generate a better solution $g_i^{(k)}$ based on the input $f_i^{(k-1)}$. Therefore, the iteration steps of the CGLS algorithm, $M_1$ is chosen to be a small integer, such as $M_1 = 3$.

To solve the subproblem in Eq. (5), let us first take functional variation of $E_1[\{f_i\}]$ with respect to $f_i(x)$. Note that the weighting factors $\omega_{i,j}^*(x,y)$ are constants defined according to the ground truth images $f_i^*(x)$ and $f_j^*(x)$. We arrive at

$$\frac{\delta E_1}{\delta f_i(x)} = 2\left(f_i - g_i^{(k)}\right) + \hat{\mu}\left(f_i - \sum_y f_{i-1}(y)\omega^*_{i,i-1}(x,y)\right)$$
$$+ \hat{\mu}\left(f_i - \sum_y f_{i+1}(y)\omega^*_{i,i+1}(x,y)\right), \quad (6)$$

where $\hat{\mu} = \mu/\beta$. By setting this variation to be zero and rearrange different terms, we obtain the optimality condition as :

$$f_i = \frac{1}{\hat{\mu}+1}g_i^{(k)} + \frac{\hat{\mu}}{2\hat{\mu}+2}\left[\sum_y f_{i-1}\omega^*_{i,i-1}(x,y) + \sum_y f_{i+1}\omega^*_{i,i+1}(x,y)\right]. \quad (7)$$

This equation leads to a Gauss-Jacobi type iteration scheme[22] for solving the problem in Eq. (5):

---
Gauss-Jacobi iteration Algorithm :

Initialize $m = 0$ and $u_i^{(0)} = g_i^{(k)}$. Do the Step 1 for $M_2$ times.
1. $u_i^{(m+1)} = \frac{1}{\hat{\mu}+1}g_i^{(k)} + \frac{\hat{\mu}}{2\hat{\mu}+2}\left[\sum_y u_{i-1}^{(m)}\omega^*_{i,i-1}(x,y) + \sum_y u_{i+1}^{(m)}\omega^*_{i,i+1}(x,y)\right]$
2. $f_i^{(k)} = u_i^{(M_2)}$

---

Again the integer $m$ here denotes iteration step. It follows from Theorem 10.1.1 in the literature[22] that such an iteration scheme converges for any $\hat{\mu} > 0$. The total number of iteration steps $M_2$ is chosen to be a small number, such as $M_2 = 1$.

Moreover, since the weighting factors $\omega^*_{i,j}(x,y)$ defined according to the ground truth images $f_i^*(x)$ and $f_j^*(y)$ are not know beforehand, we update these weights during the iteration according to the latest available images $g_i^{(k)}$ and $g_j^{(k)}$. In other words, we choose square patches from the reconstructed images obtained in the last iteration, $R_{g_i^{(k)}}(x)$ and $R_{g_j^{(k)}}(y)$, to calculate the weight $\omega_{i,j}(x,y)$ instead of $\omega^*_{i,j}(x,y)$.

$$\omega_{i,j}(x,y) = \frac{1}{Z}\exp\left[-\left\|R_{g_i^{(k)}}(x) - R_{g_j^{(k)}}(y)\right\|_2^2/h^2\right] \quad (8)$$

Additionally, a simple truncation of negative pixel values is necessary after each iteration to ensure the positivity of the reconstructed images. The TNLM algorithm can be summarized as follows:

---
TNLM Algorithm:
1. Initialize $f_i^{(0)}$ for $i = 1, \ldots, N$ to be the image reconstructed by FBP with all projections from all $N$ phases.
2. Using CGLS with initial value $\{f_i^{(k-1)}\}$ to get $\{g_i^{(k)}\}$.
3. Update weighting parameter $\omega_{i,i-1}, \omega_{i,i+1}$ using $\{g_i^{(k)}\}$.
4. Get $\{f_i^{(k)}\}$ using Gaussian-Jacobi algorithm.
5. Ensuring image positivity: $f_i^{(k)} = 0$ if $f_i^{(k)} < 0$.
6. Go back to step 2 until convergence.

---

The advantage of this TNLM algorithm is straightforward. In the $k^{th}$ iteration, the algorithm first obtains a better solution $g_i^{(k)}$ using CGLS algorithm based on the solution 190 from previous step, $f_i^{(k-1)}$. Since this step does not contain any regularization on the solution, the obtained $g_i^{(k)}$ will be contaminated by noise and various artifacts. The following TNLM step, *i.e.* step 4, updates the solution according to the Gauss-Jacobi algorithm, yielding a new solution $f_i^{(k)}$. In particular, $f_i^{(k)}$ is a weighted average of the input image $g_i^{(k)}$ and the images at neighboring phases $g_{i-1}^{(k)}$ and $g_{i+1}^{(k)}$. Moreover, the updated image $f_i^{(k)}$ pixel value at $x$ depend on $g_i^{(k)}$ in a local fashion, *i.e.* also at voxel $x$, but in a non-local fashion on the
195 neighboring phases, *i.e.* at other voxels $y$. For those non-local terms, the weight is automatically adjusted according to the similarity between the patches around $x$ in phase $i$ and the patches around $y$ in neighboring phases. As such, any features that repetitively appear in successive phases, such as true anatomical structures, are preserved during the iteration. In contrast, those features do not repeat, such as streaking artifacts, are suppressed.
200
*2.3 Implementation Issues*

One disadvantage of our TNLM reconstruction algorithm is its large computational burden of searching for similarity. During the implementation, for each pixel $x$ on image $f_i$, a
205 square patch $R_i(x)$ centering at this pixel on the image is compared with patches centered at all pixels $y$ on the neighboring phase images $f_{i\pm 1}$ for computing the weighting factor $\omega_{i,i\pm 1}(x,y)$. Suppose the patches used for computing this weighting factor are two dimensional squares with $d$ pixels in each dimension. Then the total complexity of the searching scheme between two images is in the order of $O(L^2 \times L^2 \times d^2)$, where $L$ is the size
210 of the 4D-CT slices in each dimension. However, this approach is neither computational efficient nor necessary. In fact, for a patch at location $x$ on phase $i$, the patches similar to it on neighboring phases must locate within a neighborhood of $x$ due to the finite motion range of respiratory motion. Therefore, it is adequate to restrict the search for the similar patches only within a search window. In practice, we set this search window to be a square region with $W$
215 pixels in each dimension. Since it is usually true that $W \ll L$, this searching window can reduce the computation load down to $O(L^2 \times W^2 \times d^2)$.

Another technique we used to speed up the calculation is removing the redundant calculations in computing the weighting factor. Note that the weights, $\omega_{i,j}(x,y)$ and $\omega_{j,i}(y,x)$, are actually same before normalization. Reusing this factor in the Gauss-Jacobi
220 algorithm, instead of recomputing it, can reduce the computational load by a factor of about two.

Besides, we also implement the 4DCT reconstruction algorithm on an NVIDIA Tesla C1060 card to speed up the computation. This GPU card has a total number of 240 processor cores (grouped into 30 multiprocessors with 8 cores each), each with a clock speed of 1.3
225 GHz. It is also equipped with 4 GB DDR3 memory, shared by all processor cores. We simply have each GPU thread responsible for one pixel of the CT slices. Because of the large number of GPU threads, the computation efficiency can be considerably elevated.

**3. Experimental results**
230

We tested our reconstruction algorithm on a digital NCAT phantom at a thorax region[23]. The x-ray projections were simulated using Siddon's algorithm[24] in fan-beam geometry with an arc detector of 888 units and a spacing of 1.0239mm. The source to detector distance is 949.075mm and the source to rotation center distance is 541.0mm. All of these parameters mimic the realistic configuration of a GE Lightspeed QX/I CT scanner. The gantry's rotation speed is set to be 0.5s/rotation. The period of the respiratory cycle is 4 seconds. Our 4DCT reconstruction algorithm is tested in three low dose cases: Case 1 - all projections acquired with low mAs protocol; Case 2 - undersampled projections acquired with high mAs protocol; Case 3 - undersampled projections acquired with low mAs protocol.

To simulate the noise-contaminated sinogram at low mAs situations, we add Gaussian noise signal to the noise free projection data, where the variance at a given entry *i* of the projection is taken as[25, 26]

$$\sigma_i^2 = \frac{1}{N_{0i}} \exp(\bar{p}_i) \qquad (9)$$

where $\bar{p}_i$ is the projection data value before adding noise. $N_{0i}$ represents the average photon number just before entering the patient body, which is derived from the measurements at a certain mAs level. The variance computed as such also goes into the matrix $\Sigma$ in Eq. (1).

In Case 1, we first generated 4000 noise-contaminated projections at 20mAs for ten phases and all of them are used for reconstruction. The reconstructed images at three phases, namely the end of inhale, the middle of the exhale, and the end of exhale are shown in Fig.1. The reconstruction results of the conventional FBP algorithm are also shown for comparison purpose. By visually inspecting, it is clear that our TNLM algorithm outperforms the FBP algorithm by greatly reducing the image noise. The blurring effect shown in top and bottom row images is due to the residual heart motion within the respiratory phase bin. More discussion about this effect will be given in Section 4.

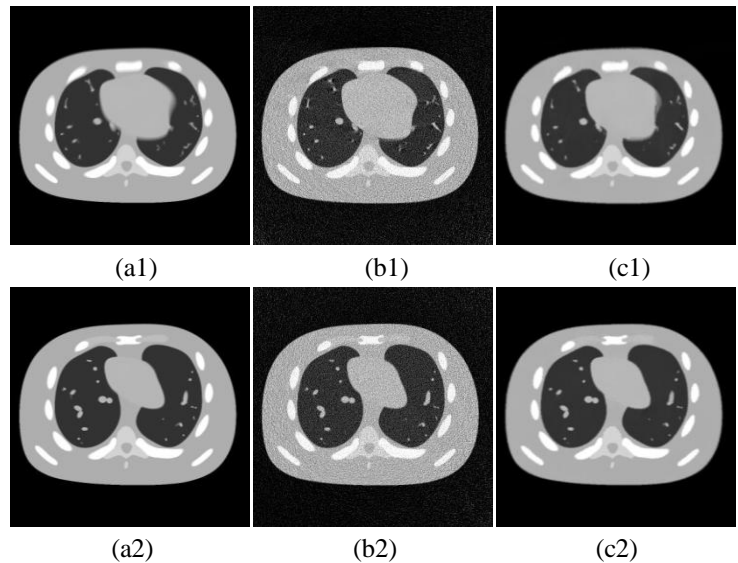

(a1)　　　　　(b1)　　　　　(c1)

(a2)　　　　　(b2)　　　　　(c2)

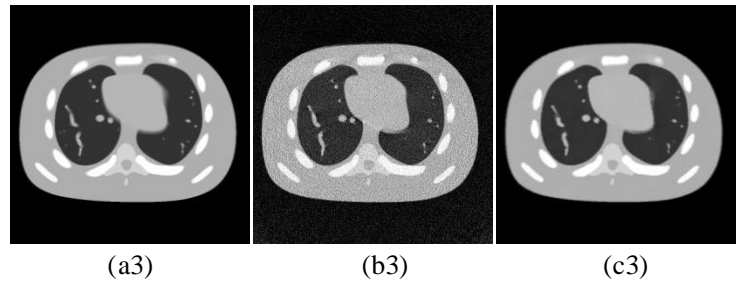

(a3) (b3) (c3)

**Figure 1.** The reconstruction 4DCT images for the 40% phase (end point of inhale) (top) and the 70% phase (mid point of exhale) (middle) and the 100% phase (end point of exhale) (bottom) with all projections at 20 mAs level (Case 1). (a) Ground truth; (b) FBP; (c) TNLM.

As for Case 2, we generated 500 projections at 100 mAs to simulate the undersampling situation with high mAs level. Case 3 is same as Case 2 except the mAs is lowered down to 20 mAs to simulate the undersampling situation with low mAs level. The results of these two cases are shown in Fig. 2 and Fig. 3, respectively. Obvious streaking artifacts can be observed in the CT images reconstructed by the FBP algorithm due to undersampling, see Fig. 2(b1~b3) and Fig. 3(b1~b3). Moreover, the FBP results of the Case 3 are noisier than that of the Case 2 due to the low mAs level. On the other hand, our TNLM algorithm can still maintain the image quality to a satisfactory level in both cases.

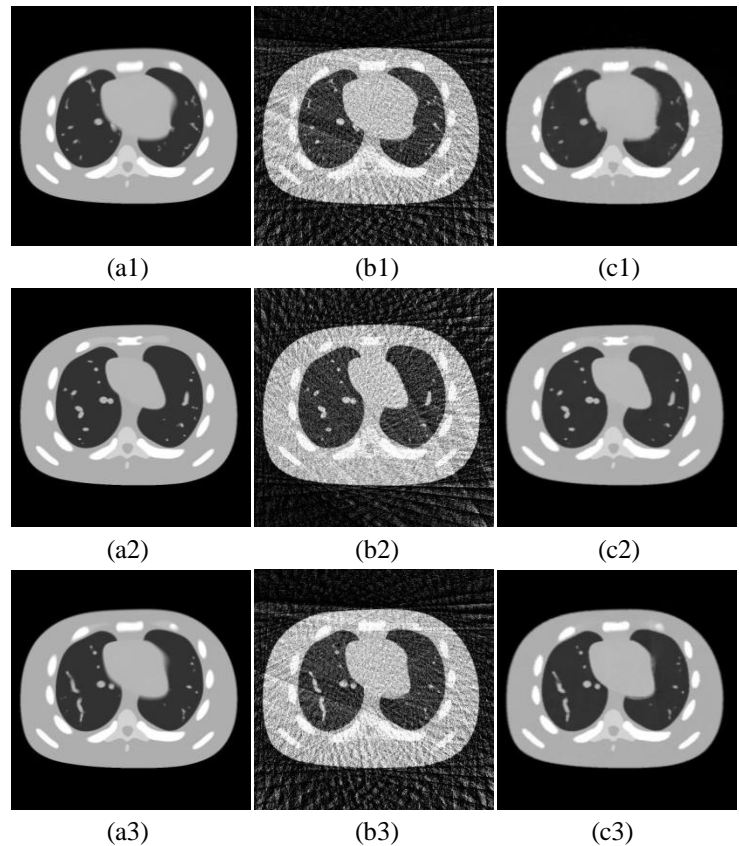

(a1) (b1) (c1)

(a2) (b2) (c2)

(a3) (b3) (c3)

**Figure 2.** The reconstruction 4DCT images for the 40% phase (end point of inhale) (top) and the 70% phase (mid point of exhale) (middle) and the 100% phase (end point of exhale) (bottom) with 500 projections at 100 mAs level (Case 2). (a) Ground truth; (b) FBP; (c) TNLM.

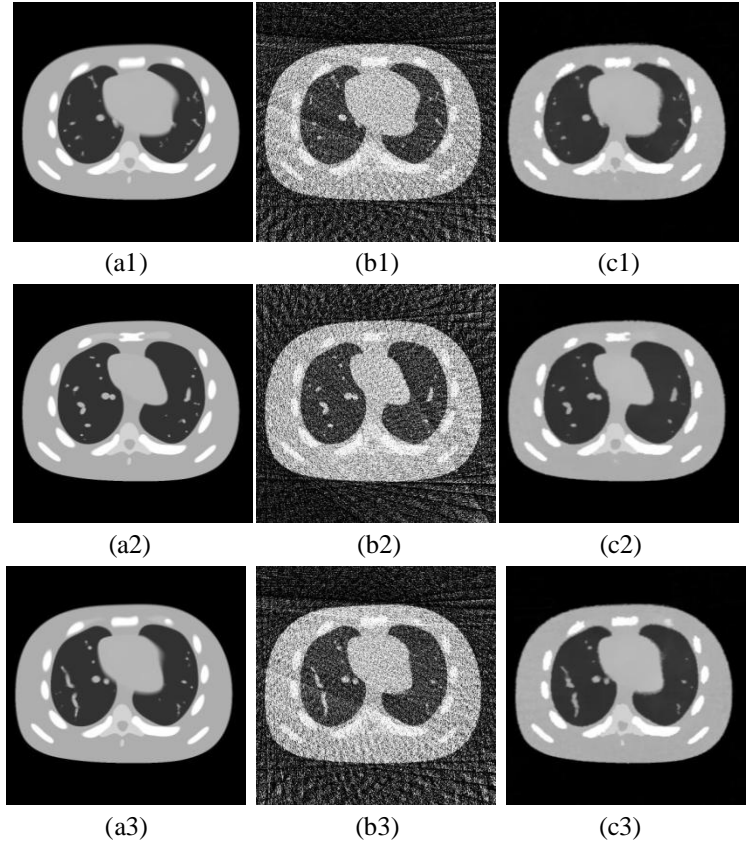

**Figure 3.** The reconstruction 4DCT images for the 40% phase (end point of inhale) (top) and the 70% phase (mid point of exhale) (middle) and the 100% phase (end point of exhale) (bottom) with 500 projections at 20 mAs level (Case 3). (a) Ground truth; (b) FBP; (c) TNLM.

To quantitatively evaluate the performance of our reconstruction algorithm in terms of maintaining image contrast and suppressing image noise, we calculate the contrast to noise ratio (CNR) and signal to noise ratio (SNR) which are defined as follows:

$$CNR = |\mu_s - \mu_b|/\sigma_b, \qquad (10)$$

$$SNR = 10\log_{10}\left\{\|f - \bar{f}\|_2^2 / \|f - f^*\|_2^2\right\}. \qquad (11)$$

The CNR is defined on a given region of interest (ROI). $\mu_s$ is the mean value of the signal for the ROI, while $\mu_b$ and $\sigma_b$ are the mean value and stand deviation of the nearby background. SNR is a quantity to measure the overall deviation of the reconstructed 4DCT image from the ground truth image. $\bar{f}$ is the mean value of an image $f$ and $f^*$ is the ground truth image.

Table 1 lists the mean values of CNRs over ten phases of the two ROIs shown in Fig. 4. ROI1 is a small high contrast structure with the background chosen as the nearby dark lung region. ROI2 is the vertebral body with the neighboring tissue as its background. In all the three cases, our TNLM algorithm has improved the CNR by a factor of 3.9-10.2 depending on the ROIs and cases compared with the conventional FBP algorithm. In particular, for the FBP algorithm, all the CNRs of ROI2 are very small and thus the low contrast structures is hardly resolved. While for our TNLM algorithm, the corresponding CNRs are much larger. The results of CNRs illustrate that our algorithm is superior to the FBP algorithm in maintaining good contrast under low dose contexts. Table 2 lists the mean SNRs over ten phases, in which our TNLM algorithm has improved the SNRs by a factor of 2.1-5.9, indicating that our algorithm outperforms the FBP algorithm in effectively suppressing the image noise at low dose situation.

|      | Case1 |      | Case2 |      | Case3 |      |
|------|-------|------|-------|------|-------|------|
|      | ROI1  | ROI2 | ROI1  | ROI2 | ROI1  | ROI2 |
| FBP  | 6.83  | 1.80 | 4.09  | 1.13 | 2.56  | 0.69 |
| TNLM | 28.64 | 12.15| 16.04 | 7.53 | 19.53 | 7.07 |

**Table 1.** Mean CNR values over ten phases of the two ROIs in Fig. 4.

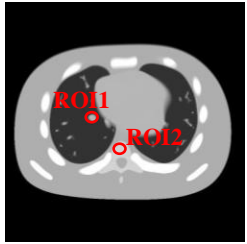

|      | Case1 | Case2 | Case3 |
|------|-------|-------|-------|
| FBP  | 10.49 | 6.03  | 3.39  |
| TNLM | 22.13 | 21.12 | 20.28 |

**Table 2.** Mean SNR values over ten phases.

**Figure 4.** Two ROIs used for CNR calculation.

**4. Discussion and Conclusions**

In this paper, we have presented a novel iterative 4DCT reconstruction algorithm via temporal regulariation. The 4DCT images of different phases are reconstructed simultaneously by minimizing an energy function consisting of a data fidelity term and a temporal regularization term between every two neighboring phases. A temporal non-local means method is employed to take the temporal correlation of the 4DCT images into account. We utilized a modified forward-backward algorithm to perform the optimization. The iterative reconstruction algorithm is implemented on a GPU platform to improve its efficiency. Our algorithm is tested on a digital NCAT phantom under low dose context by lowering the mAs level or/and decreasing the number of projections. The experimental results indicate that our algorithm performs much better than the conventional FBP algorithm in effectively reducing the image artifacts due to undersampling and suppressing the image noise due to the low mAs level. Specifically, the contrast-to-noise ratio has been improved by a factor of 3.9-10.2 and the signal-to-noise ratio has been improved by a factor of 2.1-5.9, depending on the cases. The total reconstruction time ranges from 40 to 90 seconds on a NVIDIA Tesla C1060 card (NVIDIA, Santa Clara, CA) for ten phases at a transverse slice. This reconstruction time may not meet the requirement for some clinical applications. Yet, the efficiency of our algorithm could be further increased by using some advanced technologies such as multi-grid algorithm and multiple GPUs.

The blurred heart edges in some panels of Figs. 1-3 are caused by the residual heart motion within the respiratory phase bin. The breathing period used in this work is 4 sec. Then each of 10 breathing phase bin covers 0.4 sec, within which the respiratory motion is negligible but the heart motion is not. For example, we observe apparent blurring effect in Fig. 1(c1) and Fig. 1(c3), but not Fig. 1(c2), this is because the heart in Fig. 1(c1) and Fig. 1(c3) is in the middle of systole/diastole, while in Fig. 1(c2) it happens to be at the end of systole. The the blurring effect also presents in the ground truth images, as shown in Fig. 1(a1) and Fig. 1(a3), because we average over 400 images within a given breathing phase bin to produce the ground truth CT image at that phase. The blurring effect due to the residual heart motion also exists in the FBP images, which, however, is less visible due to the image noise and the streak artifact.

In this feasibility study we have only quantitatively evaluated CNR and SNR. Other quantitative measures, such as spatial resolution of the images, are not considered in the current work. It is found that the spatial resolution is degraded to a certain extent with the TNLM method, especially at the undersampling situation in Case 2 and Case 3. We plan to conduct much more systematical and quantitative evaluations using a large set of representative patient images and a set of clinically relevant metrics in our future work.

**Acknowledgements**

This work is supported in part by the University of California Lab Fees Research Program. We would like to thank NVIDIA for providing GPU cards for this project.